%% file: main.tex
\documentclass[a4paper,fleqn]{cas-sc}
\usepackage{lineno} 

\usepackage[authoryear]{natbib}

\def\tsc#1{\csdef{#1}{\textsc{\lowercase{#1}}\xspace}}
\tsc{WGM}
\tsc{QE}

\begin{document}
\let\WriteBookmarks\relax
\def\floatpagepagefraction{1}
\def\textpagefraction{.001}

\shorttitle{Femtoscopy Measurement with S$\pi$RIT TPC in Radioactive Beam Heavy-ion Collisions}    

\shortauthors{Y. J. Wang et~al.}  

\title [mode = title]{Femtoscopy Measurement with S$\pi$RIT TPC in Radioactive Beam Heavy-ion Collisions}  



%

\include{authors}


\begin{abstract}
Femtoscopy is a powerful tool for exploring the dynamic emitting structure in heavy-ion collisions, while radioactive beam heavy-ion collisions enable the investigation of nuclear matter under extreme isospin conditions. Here, we successfully perform femtoscopy measurements using the S$\pi$RIT Time Projection Chamber (TPC). A dedicated correction scheme for track merging and splitting is proposed, which is well applicable to rectangular TPCs housed inside dipole magnets and effectively improves the reconstructed correlation functions at small relative momenta. Focusing on the proton–proton (p-p) correlation function in the 270~{AMeV} $^{132}\text{Sn}+^{124}\text{Sn}$ system, we successfully apply the track merging and splitting correction; additionally, the TPC angular acceptance exhibits a negligible impact on the correlation function. A systematic uncertainty quantification framework is established. The experimental results of the p-p correlation function confirm the feasibility of the S$\pi$RIT TPC for femtoscopy measurements and provide technical support for high-precision femtoscopy studies using rectangular TPCs in radioactive beam heavy-ion collisions.
\end{abstract}





\maketitle


\section{Introduction}\label{Introduction}

Heavy-ion collisions serve as a unique experimental platform to explore the properties of nuclear matter under extreme conditions, such as  {net-}baryon density, high temperature, and large isospin asymmetry, which are crucial for understanding the equation of state (EoS) of nuclear matter and the astrophysical processes in neutron stars and neutron star mergers \cite{li2008recent, Huth:2021bsp, Tsang:2023vhh, Xiao:2008vm}. Among the various techniques employed to probe the spatio-temporal characteristics of the collision-generated system, femtoscopy, based on two-particle intensity interferometry, stands out as a powerful tool \cite{Tam:2025mkk, Xu:2024dnd, SRIT:2026qkv, Verde:2003cx, Zhang:2025yqq}. It enables the measurement of the particle-emitting source on the femtometer scale (10$^{-15}$ meters), providing invaluable insights into the dynamics of heavy-ion collisions \cite{Wang:2021mrv, Duan:2026cst}.

Femtoscopy measurements rely on the analysis of momentum space correlations between pairs of emitted particles, which are governed by the emitting source, quantum statistics and final-state interactions, including strong and Coulomb interactions. Traditional femtoscopic studies have primarily focused on stable beams (e.g., Ar+Au, Sn+Sn), yielding significant progress in particle emission sequence \cite{Wang:2021mrv} and neutron-neutron interactions \cite{Si:2025eou}. However, these measurements are limited in their ability to explore the isospin dependence of nuclear matter properties, a critical gap that can be addressed by employing radioactive isotope beams  with extreme neutron-to-proton (N/Z) ratios \cite{SpiRIT:2021gtq}.

Radioactive beam heavy-ion collisions have opened a new frontier in nuclear physics. They enable the study of nuclear matter at large isospin asymmetries, regimes unattainable with stable beams and highly representative of the environments inside neutron stars \cite{SpiRIT:2021och, SpiRIT:2022sqt, SpiRIT:2023htl}. The study of correlations between particles emitted in radioactive beam heavy-ion collisions provides enhanced sensitivity to the isospin degree of freedom, offering a unique opportunity to constrain the density dependence of the symmetry energy term in the nuclear EoS, a long-standing goal in both nuclear physics and astrophysics. However, femtoscopy measurements with radioactive beams pose significant experimental challenges: the low intensity of radioactive beams, the copious production of light charged particles (LCPs) and pions in central collisions, and the need for high-precision tracking and particle identification (PID) over a large solid angle to capture the full momentum range of correlated particle pairs.

To address these challenges, the Superconducting Analyzer for Multi-particles from Radioisotope (SAMURAI) Pion-Reconstruction and Ion-Tracker (S$\pi$RIT) Time Projection Chamber (TPC) was specifically designed and constructed for experiments at the RIKEN Radioactive Isotope Beam Factory (RIBF) \cite{SpiRIT:2014yhq,SpiRIT:2016vpd,SpiRIT:2019jjg,Barney:2020mxk,Lasko:2016igj,Isobe:2018udc}. As a high-performance tracking detector, the S$\pi$RIT TPC offers exceptional capabilities tailored to radioactive beam heavy-ion collision experiments. Moreover, the S$\pi$RIT TPC is optimized to work in conjunction with the SAMURAI spectrometer. This design ensures efficient reconstruction of particle tracks with the SAMURAI dipole magnet, making it suitable for measuring the momentum distributions of particles emitted in heavy-ion collisions, expanding its versatility for femtoscopic studies.

Despite the potential of the S$\pi$RIT TPC for femtoscopy measurements, several technical and analytical challenges persist. These include correcting the track merging and track splitting effect on correlation statistics, accounting for  the influence of geometrical acceptance effect, and estimating the systematic uncertainty. { Addressing these challenges enables high-precision measurements of the two-proton correlation function in the low relative momentum region using the S$\pi$RIT TPC. Conventional low-energy femtoscopy experiments relying on discrete solid-state detectors generally cannot access the relative momentum region below $17$-$18$ MeV/c with sufficient resolution \cite{Wang:2021jgu, Valdre:2018dni}. This low relative momentum region is physically crucial, as it is dominated by Coulomb interactions between proton pairs emitted from long-lived particle sources. An inaccurate description of the low relative momentum correlation function in source fitting directly degrades the precision of extracted source sizes and undermines the reliability of femtoscopy analyses. The high-quality low relative momentum data acquired with the S$\pi$RIT TPC allows the extraction of source sizes with sufficiently small uncertainties, enabling quantitative investigations of the N/Z dependence of particle-emitting sources. This will also lay a robust methodological foundation for future high‑precision radioactive beam femtoscopy measurements using rectangular TPC detectors.}

In this paper, we present a comprehensive study of femtoscopy measurements using the S$\pi$RIT TPC in radioactive beam heavy-ion collisions. We briefly introduce the experimental setup, including the S$\pi$RIT TPC’s performance characteristics in terms of particle identification and multiplicity distribution. We focus on the measurement of proton-proton correlation function in 270~{AMeV} $^{132}$Sn + $^{124}$Sn. A correction scheme for track merging and track splitting is proposed to improve the performance of the experimental correlation function at small relative momenta. The effect of geometrical acceptance is explored using different geometrical acceptance cuts, and the estimation of systematic uncertainty is discussed in detail.
Furthermore, this work establishes the S$\pi$RIT TPC as a robust tool for future femtoscopic studies with exotic radioactive beams, paving the way for investigations of extreme nuclear matter conditions relevant to astrophysics and fundamental nuclear physics.

\section{Brief description of S$\pi$RIT experiment}\label{Exp}

\begin{figure*}
	\centering
	\includegraphics[width=.9\textwidth]{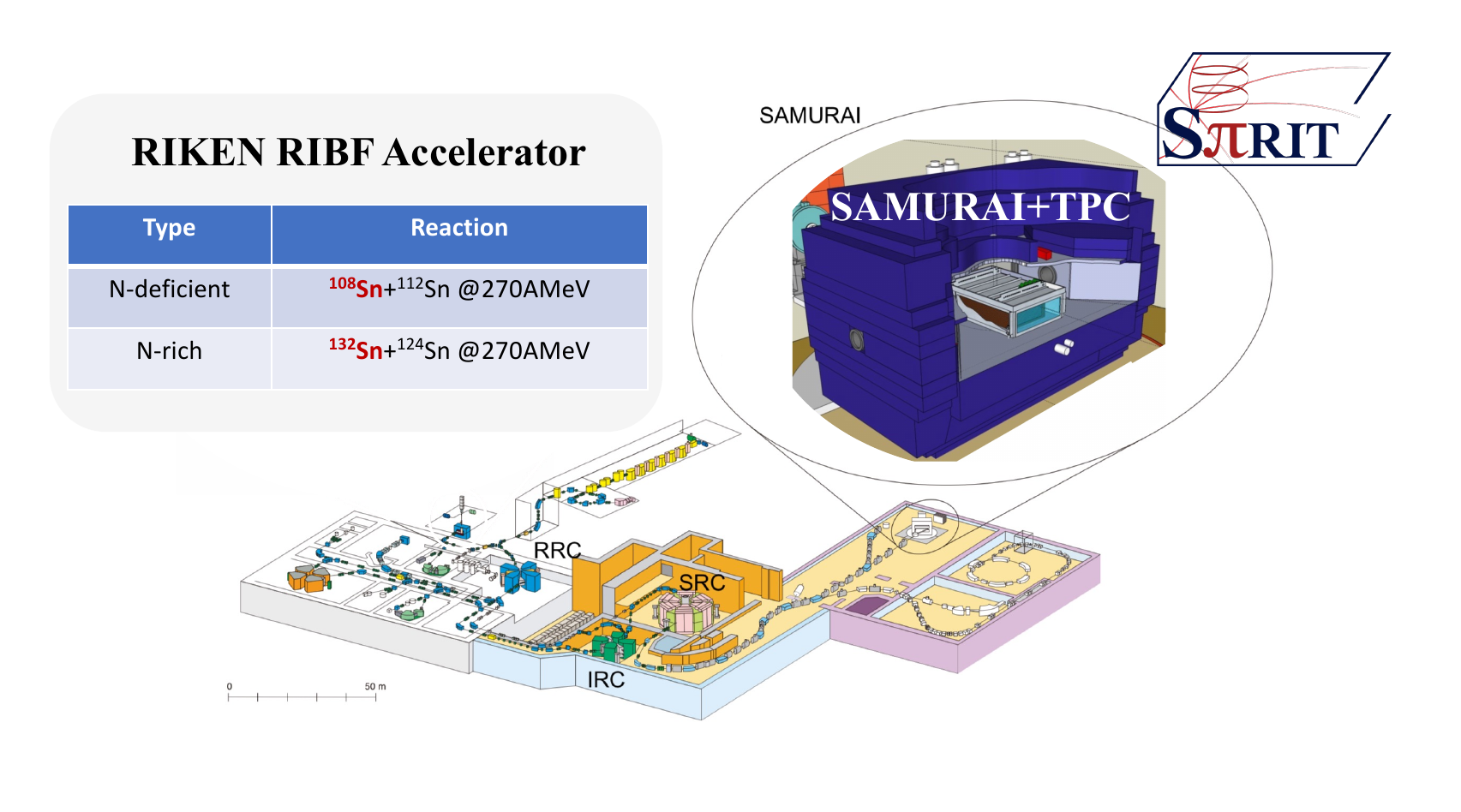}
	\caption{Layout diagram of the RIKEN radioactive isotope factory accelerator and schematic diagram of the S$\pi$RIT experimental setup \cite{SpiRIT:2014yhq,SpiRIT:2016vpd,SpiRIT:2019jjg,Barney:2020mxk}. The SAMURAI terminal, where the S$\pi$RIT experiment is installed, is explicitly marked. { The S$\pi$RIT TPC \cite{SpiRIT:2014yhq,SpiRIT:2016vpd,SpiRIT:2019jjg,Barney:2020mxk,Lasko:2016igj,Isobe:2018udc}, mounted inside the SAMURAI superconducting dipole magnet \cite{OTSU2016175},  was employed to measure the trajectories of charged particles.}  A table compiling the radioactive beam heavy-ion collision systems is provided.}
	\label{fig1}
\end{figure*}

This experiment was performed at the SAMURAI experimental area of the RIBF at the RIKEN Nishina Center for Accelerator-Based Science in Japan, operated by the S$\pi$RIT Collaboration (Fig. \ref{fig1}). The experimental data analyzed in this work were obtained from the S$\pi$RIT experiment conducted in 2016. The experiment employed $^{108}$Sn and $^{132}$Sn beams at an energy of 270~{AMeV}, bombarding $^{112}$Sn and $^{124}$Sn targets with isotopic abundances greater than 95$\%$, with areal densities of 561~mg/cm$^{2}$ and 608~mg/cm$^{2}$, respectively.

The trajectories of charged particles were detected by the S$\pi$RIT TPC \cite{SpiRIT:2014yhq,SpiRIT:2016vpd,SpiRIT:2019jjg,Barney:2020mxk,Lasko:2016igj,Isobe:2018udc}, which was installed inside the SAMURAI superconducting dipole magnet \cite{OTSU2016175}. { The TPC features a 50.5 cm drift length and an \(86.4\ \mathrm{cm} \times 134.4\ \mathrm{cm}\) pad plane. Its field cage is operated with P10 gas (\(90\%\) Ar + \(10\%\) \(\mathrm{CH_4}\)). The pad plane comprises 108 rows along the z‑axis and 112 columns along the x‑axis. Each pad is \(7.5\ \mathrm{mm}\,(x) \times 11.5\ \mathrm{mm}\,(z)\), with \(0.5\ \mathrm{mm}\) gaps between adjacent pads \cite{Barney:2020mxk}. Signals from the pad plane are read out using the Generic Electronics for TPC (GET) system \cite{Isobe:2018udc}.} The magnetic field strength was set to 0.5 T during the experiment, which could simultaneously meet the measurement requirements for light charged particles and $\pi$ mesons. 

\begin{figure}
	\centering
	\includegraphics[width=.8\textwidth]{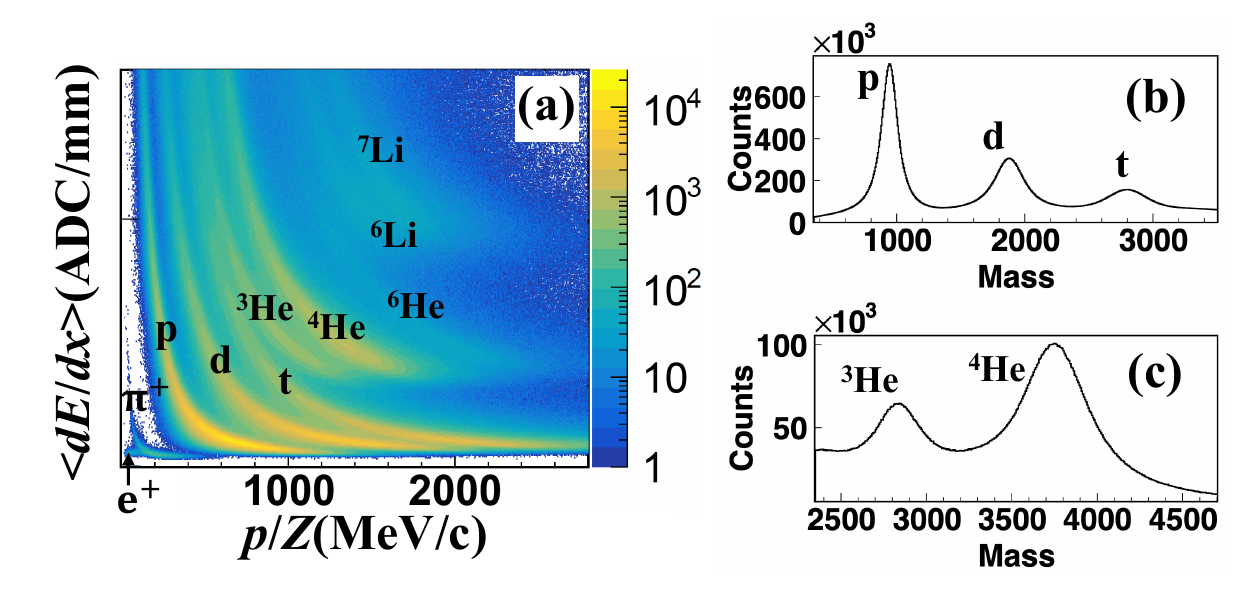}
	\caption{Particle identification performance of the S$\pi$RIT TPC in 270~{AMeV} $^{132}$Sn + $^{124}$Sn (a). Mass numbers of hydrogen isotopes (proton, deuteron, triton) with charge number $Z=1$ (b) and helium isotopes ($^{3}$He, $^{4}$He) with charge number $Z=2$ (c) are extracted respectively via the Bethe–Bloch formula.}
	\label{fig2}
\end{figure}

The space charge effect was considered in { S$\pi$RITROOT (an analysis software framework for the S$\pi$RIT experiment) \cite{SpiRIT:2020gce, SpiRITGithub},} the interference of beam-induced electric field distortions on particle track measurements has been effectively eliminated \cite{SpRIT:2019xqz}. Furthermore, the emission vertex position and magnetic rigidity ($B\rho = p/Z$) of particles are obtained by fitting the particle tracks, { where GENFIT \cite{Rauch:2014wta} is employed for momentum reconstruction and RAVE \cite{Waltenberger:2011zz} for vertex reconstruction within in S$\pi$RITROOT software framework. Energy loss effects inside the target, entrance windows, and detector gas are fully accounted for during particle‑momentum reconstruction \cite{Barney:2020mxk}.} Combining the average energy loss ($<dE/dx>$) and magnetic rigidity ($B\rho = p/Z$) information of the TPC, the identification of different particles is ultimately achieved. Figure \ref{fig2} (a) demonstrates the particle identification performance of the S$\pi$RIT TPC in 270~{AMeV} $^{132}$Sn + $^{124}$Sn. Furthermore, the mass numbers of hydrogen isotopes (proton, deuteron, triton) with charge number $Z=1$ and helium isotopes ($^{3}$He, $^{4}$He) with charge number $Z=2$ are extracted respectively via the Bethe–Bloch formula, providing key data support for subsequent physical analysis (Fig. \ref{fig2} (b-c)).

\begin{figure}
	\centering
	\includegraphics[width=.4\textwidth]{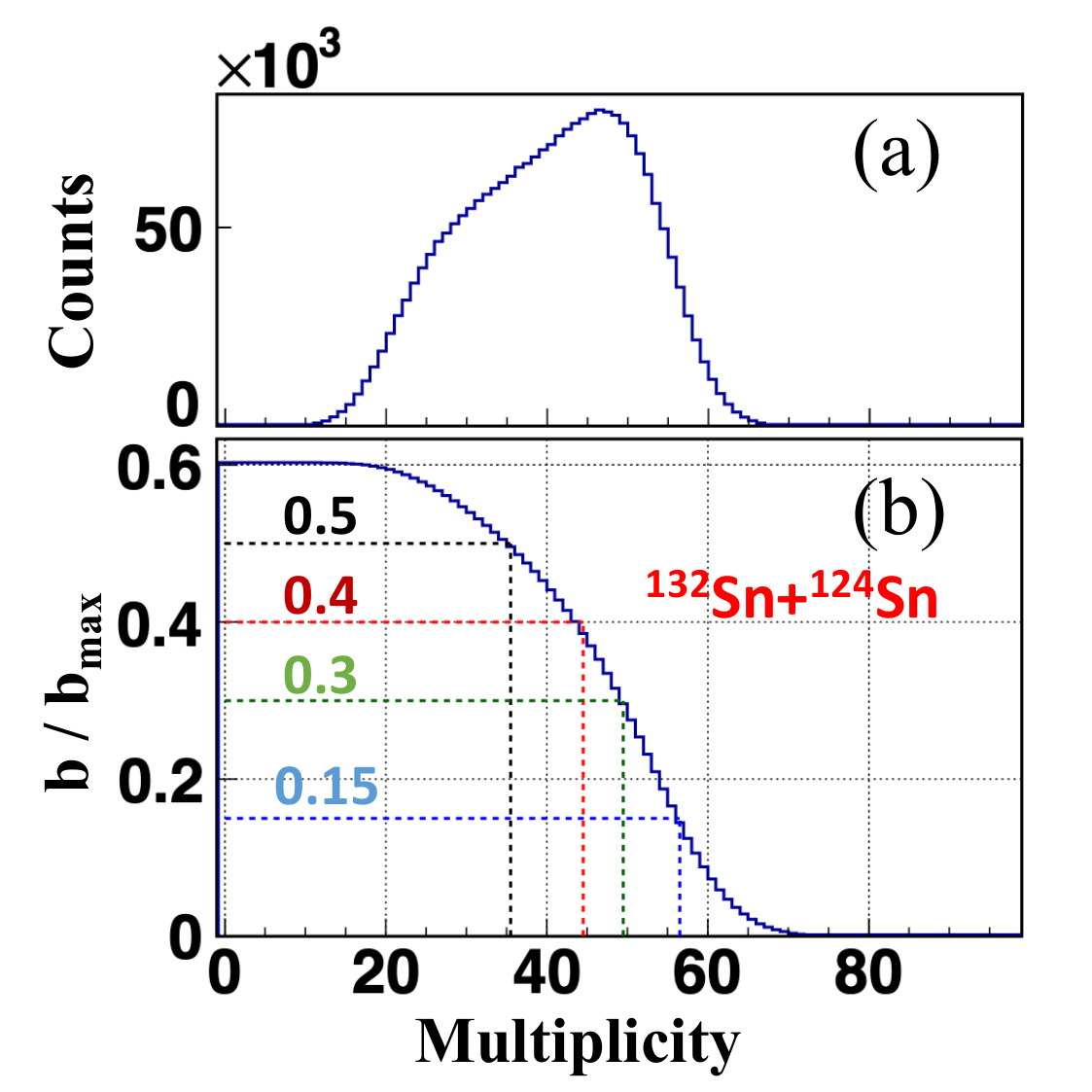}
	\caption{The multiplicity distribution of the charged particles detected with S$\pi$RIT TPC for $^{132}\text{Sn}+^{124}\text{Sn}$ reaction system (a). The mapping relation between  $b/b_{max}$ and the particle multiplicity for $^{132}\text{Sn}+^{124}\text{Sn}$ reaction (b).}
	\label{fig3}
\end{figure}

The centrality was determined with the multiplicity of the charged particles detected with S$\pi$RIT TPC. The multiplicity distribution of the charged particles is shown in Fig. \ref{fig3} (a), which is filtered by the experimental trigger condition. The ratio between the impact parameter corresponding to the experimental trigger and the maximum impact parameter ($b_{\text{trigger}}/b_{\text{max}}$) can be determined by the  formula:

\begin{equation}
\frac{b_{\text{trigger}}}{b_{\text{max}}}=\sqrt{\frac{\sigma^{\text{trigger}}}{\sigma^{\text{max}}}} = \sqrt{\frac{N_{\text{trigger}}}{N_{\text{max}}}} ,
\label{eq_b_trigger}
\end{equation}

where $\sigma^{\text{trigger}}$ and $\sigma^{\text{max}}$ denote the cross section corresponding to the experimental trigger and the maximum cross section, respectively. $N_{\text{trigger}}$ and $N_{\text{max}}$ represent the number of valid events for the experimental trigger and the maximum number of valid events, respectively. Considering the hard sphere model ($b_{\text{max}}=1.15\left(A_{\text{P}}^{1/3}+A_{\text{T}}^{1/3}\right)$) and the dead time of the acquisition system, the calculated ratio is the $b_{\text{trigger}}/b_{\text{max}}=0.603$ for the $^{132}\text{Sn}+^{124}\text{Sn}$ reaction system.

The mapping between particle multiplicity and impact parameter is established using Eq.\ref{eq_Multipliciy}.

\begin{equation}
b_{mc}(m_{c})=b_{\text{trigger}}\left(\sum_{M_{c}=m_{c}}^{\infty} \frac{dP(M_{c})}{dM_{c}}\right)^{1/2}
\label{eq_Multipliciy}
\end{equation}

Here, $b_{mc}(m_{c})$ refers to the impact parameter matched to particle multiplicity $m_{c}$, and  $dP(M_{c})/dM_{c}$ describes the normalized probability distribution of multiplicity $M_{c}$.  For the $^{132}\text{Sn}+^{124}\text{Sn}$ reaction, the mapping  between  $b/b_{max}$ and multiplicity is displayed in Fig. \ref{fig3}(b), which acts as the reference for classifying collision centrality.

\section{Femtoscopy}\label{HBT}

Femtoscopy, also known as intensity interferometry, is an essential technique for heavy-ion collision studies \cite{Lisa:2005dd}. Its core principle is to construct the two-particle correlation function via measurements of two-particle coincidence events. Due to final-state interactions between the emitted particle pair, the experimentally derived correlation function enables reconstruction of the spatio-temporal characteristics of the particle emission source \cite{Wang:2021mrv,Xu:2024dnd}, as well as extraction of interparticle interaction parameters \cite{STAR:2015kha,Si:2025eou}.

Experimentally, the particle correlation function in femtoscopy is constructed by dividing the relative momentum spectrum from coincidence events by the uncorrelated event spectrum obtained via the mixed-event method \cite{Lisa:1991zz,Kopylov:1974th}, followed by normalization at large relative momenta. The explicit experimental definition reads:
\begin{equation}
C_{\rm exp}(k^{*})=A\frac{N^{\rm same}(k^{*})}{N^{\rm mix}(k^{*})}
\label{eq_exp_HBT}
\end{equation}

where $k^{*}=\left|\mathbf{p}_{1}-\mathbf{p}_{2}\right|/2$ is the relative momentum of the two particles in their pair rest frame. $N^{\rm same}(k^{*})$ denotes the relative momentum spectrum constructed from correlated coincidence events, and $N^{\rm mix}(k^{*})$ represents the uncorrelated background spectrum produced by the mixed-event method \cite{Lisa:1991zz,Kopylov:1974th}. The coefficient $A$ is a normalization constant that constrains the correlation function to unity in the large $k^{*}$.

Theoretically, by incorporating the two-particle scattering wave function, femtoscopy establishes a quantitative connection between the correlation function and the spatio-temporal profile of the particle emission source. The theoretical correlation function is formulated as:
\begin{equation}
C({k^{*}}) = \int d^3r S({r}) |\psi_{{k^{*}}}({r})|^2
\end{equation}

\begin{figure}
	\centering
	\includegraphics[width=.6\textwidth]{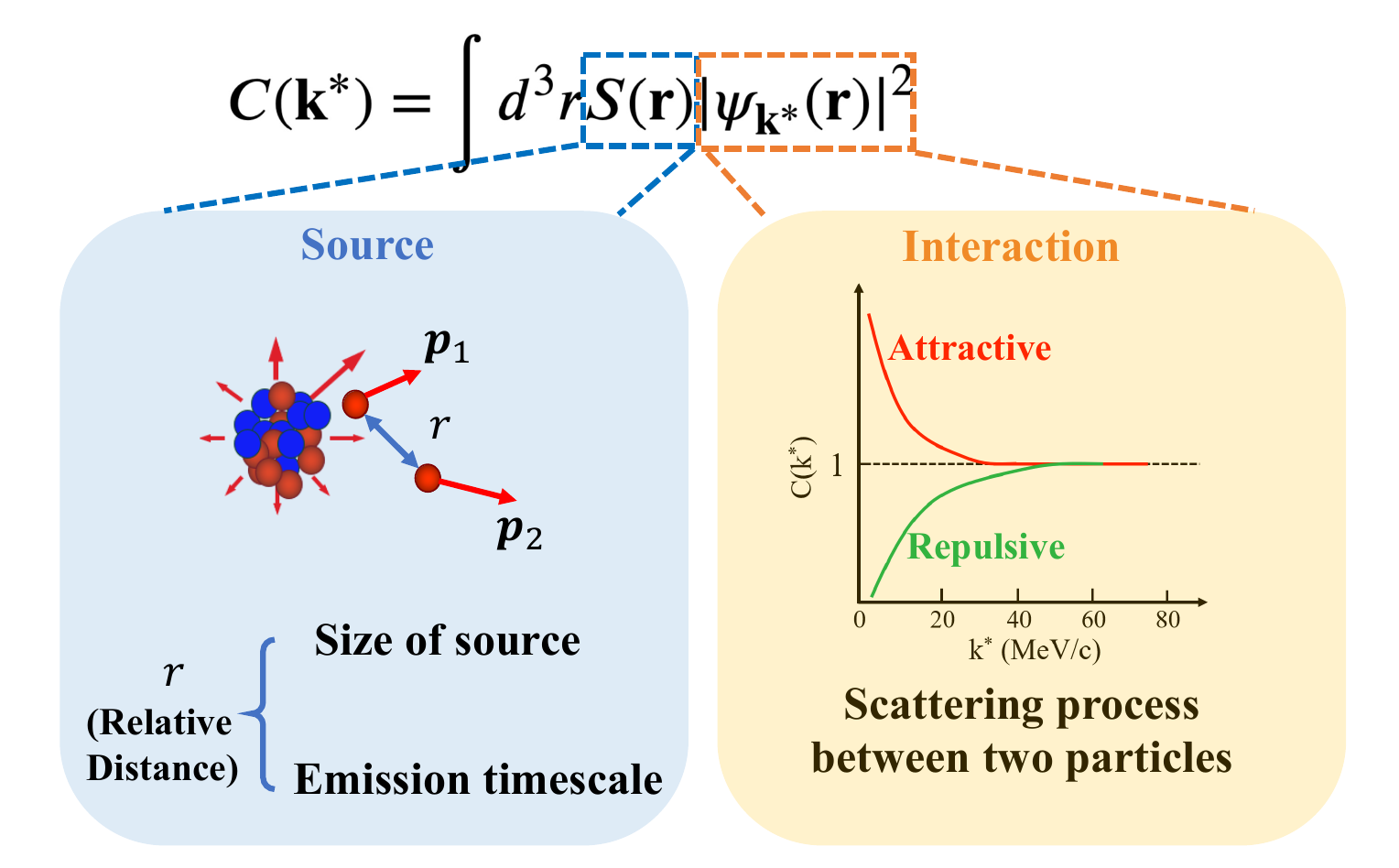}
	\caption{Schematic of the theoretical principle for particle correlation function.}
	\label{fig4}
\end{figure}

Here, $S({r})$ is the space-time distribution function of the particle emission source, and $\psi_{{k^{*}}}({r})$ is the two-particle scattering wave function accounting for final-state interactions. By fitting theoretical calculations to experimental measurements, { the space-time information of particle emission \cite{ALICE:2026tqh,ALICE:2025wuy} and the interaction between particles \cite{ALICE:2026xux,ALICE:2025kma} can be extracted} (Fig. \ref{fig4}).

\section{Track merging and track splitting effect}\label{TMTS}

When analyzing particle correlation functions using track detectors such as the TPC, track merging and track splitting are two dominant systematic effects that affect the measurement precision of correlation functions.
Track merging occurs when the tracks produced by two particles in the detector are too close to each other and are misidentified as a single track. This leads to a relative depletion of events in the low relative-momentum region of the correlation function.
In contrast, track splitting refers to the case where the track of a single particle is incorrectly reconstructed as multiple separate tracks. This spurious reconstruction introduces fake particle pairs into the correlation function, generally resulting in an artificial excess of events at low relative momentum.
To accurately measure the particle correlation function, these two effects must be effectively corrected. In this work, we propose a track merging and splitting correction methods for rectangular track detectors, based on the method from STAR experiment \cite{STAR:2024zvj,STAR:2004qya}, combined with the specific configuration of the S$\pi$RIT TPC. We perform the corresponding correction on the TPC data to improve the accuracy of the extracted correlation function.

\begin{figure}
	\centering
	\includegraphics[width=.8\textwidth]{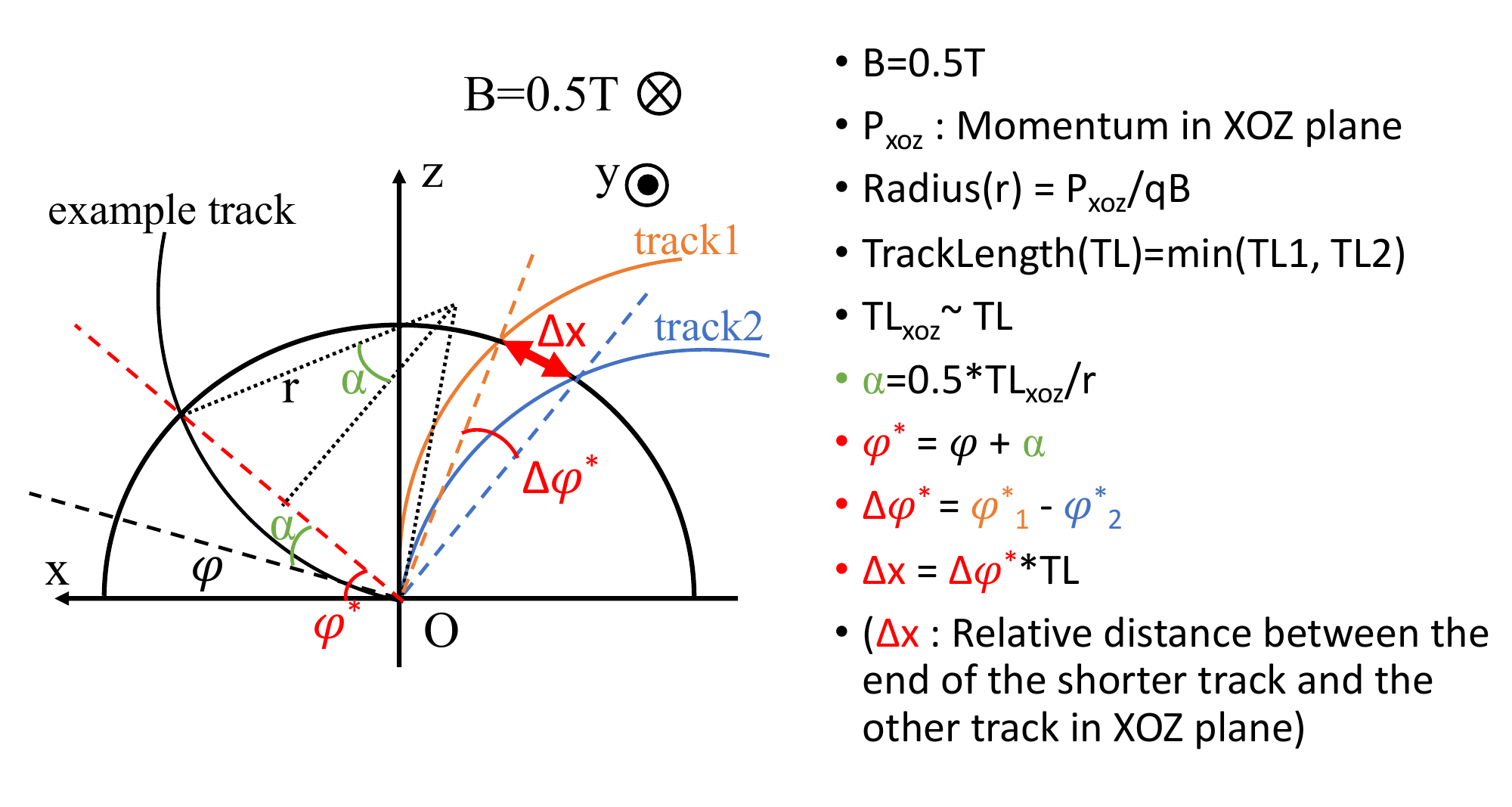}
	\caption{Schematic diagram of track merging and track splitting correction of particle tracks on the XOZ plane of TPC (plane perpendicular to the magnetic field direction).}
	\label{fig5}
\end{figure}

Figure \ref{fig5} shows a schematic diagram of the corrections for track merging and track splitting of particle trajectories in the XOZ plane of the TPC (the plane perpendicular to the magnetic field direction). The detailed correction procedure is described as follows. First, the curvature radius $r$ of the particle track in the XOZ plane is calculated using the particle momentum, charge, and magnetic field strength in this plane. Subsequently, the shorter one of the two analyzed particle tracks is selected, and its length is defined as the track length (TL) of the particle evolving in the magnetic field. Here, the actual particle track length in the XOZ plane is approximated to be consistent with TL.
Furthermore, with the track length TL and curvature radius $r$, the deflection angle $\alpha$ relative to the X-axis after the particle propagates over the distance TL is determined. Combined with the angle $\varphi$ between the initial state of the particle at the target position and the X-axis, the directional angle $\varphi^{*}$ relative to the X-axis after propagating over the distance TL is obtained. Finally, by comparing the directional angles $\varphi^{*}$ of the two particle tracks after propagating over the distance TL, the relative angular difference $\Delta\varphi^{*}$ is calculated. Given that the two tracks share a common starting point,
we combine the trajectory length TL with $\Delta\varphi^{*}$ to calculate the relative distance $\Delta x$ between the two trajectories at the shorter trajectory endpoint. This quantity describes the spatial separation of the two tracks in the 
XOZ plane.

\begin{figure}
	\centering
	\includegraphics[width=.8\textwidth]{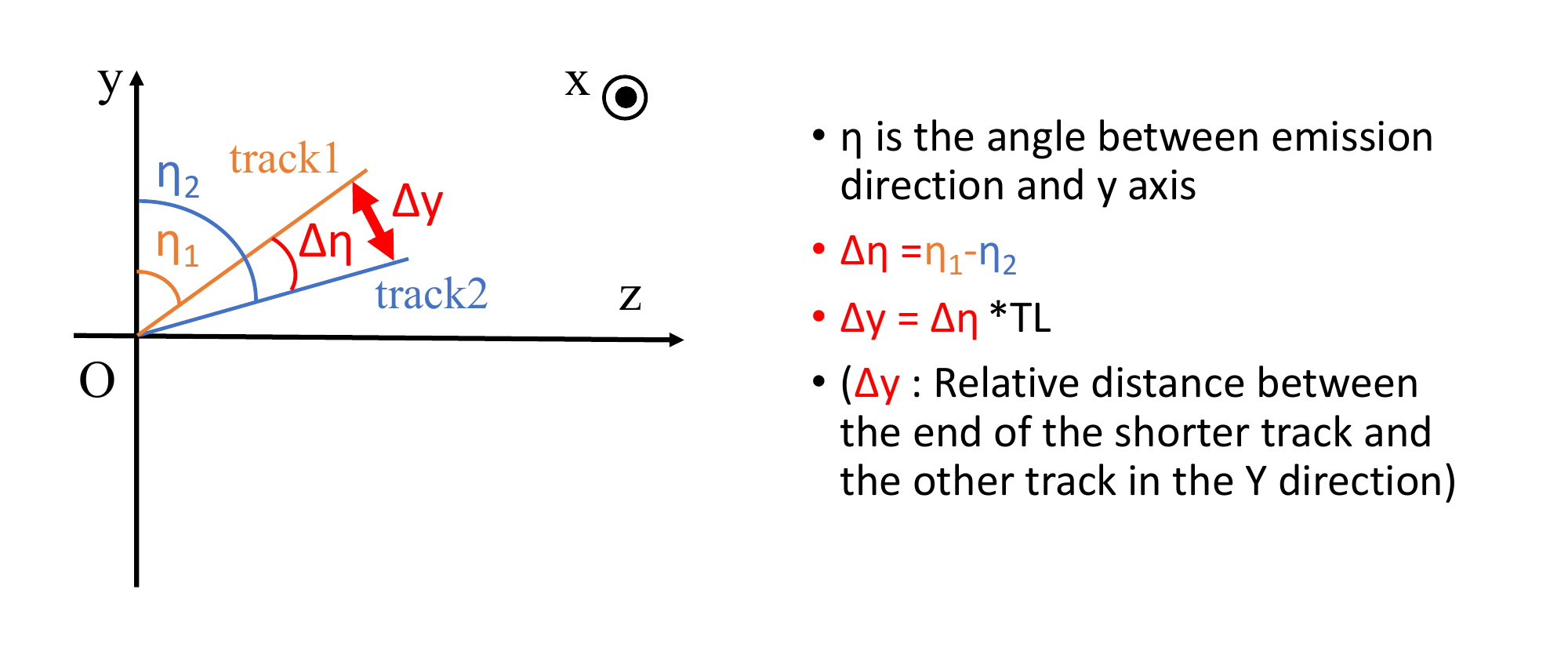}
	\caption{Schematic diagram of track merging and track splitting correction of particle tracks in the Y-axis direction (parallel to the magnetic field) of TPC.}
	\label{fig6}
\end{figure}

Figure \ref{fig6} presents a schematic of track merging and splitting corrections for particle trajectories along the TPC Y-axis, which is parallel to the magnetic field. Here, $\eta$ is the angle between an individual track and the Y-axis, and $\Delta\eta$ represents the angular separation between the two tracks under analysis. To quantify the Y-axis resolution of two tracks originating from a common vertex, we use the relative distance $\Delta y$ between them at the endpoint of the shorter track as the evaluation metric.

\begin{figure}
	\centering
	\includegraphics[width=.8\textwidth]{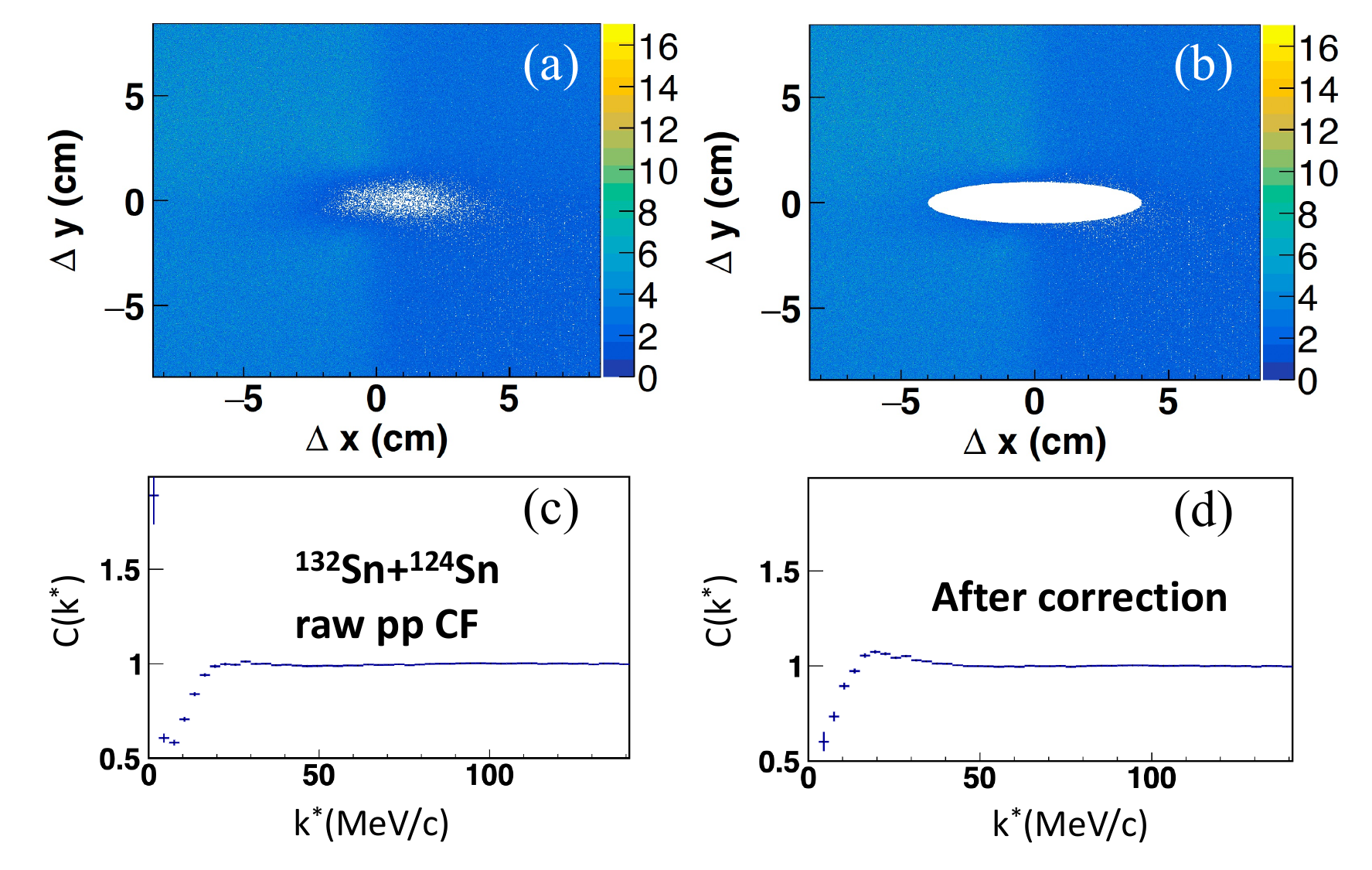}
	\caption{Distributions of track separation distance ($\Delta x - \Delta y$) in the XOZ plane and Y-axis direction, together with the proton-proton correlation functions, before and after track merging and track splitting corrections for the $^{132}\text{Sn}+^{124}\text{Sn}$ reaction system. Panel (a) shows the $\Delta x - \Delta y$ distribution before correction, and panel (c) presents the corresponding proton-proton correlation function before correction; panel (b) displays the corrected $\Delta x - \Delta y$ distribution, while panel (d) gives the proton-proton correlation function after correction.}
	\label{fig7}
\end{figure}

Figure \ref{fig7} compares the track separation distance distributions in the XOZ plane and 
Y-axis direction ($\Delta x - \Delta y$), as well as the corresponding proton-proton correlation functions, before and after track merging and track splitting corrections for the $^{132}\text{Sn}+^{124}\text{Sn}$ reaction system.
As demonstrated in Fig. \ref{fig7}(a), due to the track merging and splitting effects in the track reconstruction procedure, non-uniform track detection efficiency is observed in the region where $\Delta x$ and $\Delta y$ approach zero prior to correction. Taking into account the features of the experimental data distribution and the track resolution performance of the TPC in the XOZ plane and along the Y-axis, an elliptical region selection criterion is adopted in this analysis, as shown in Fig. \ref{fig7}(b). The ellipse is defined with semi-axes of 4 cm along $\Delta x$ and 1 cm along  $\Delta y$. Rejecting all tracks falling inside this elliptical region effectively removes reconstruction biases induced by track merging and splitting.
A comparison of the proton-proton correlation functions before and after correction in Fig. \ref{fig7}(c) and (d) indicates that the present correction remarkably optimizes the distribution shape of the proton-proton correlation function in the low relative-momentum region, which provides an essential foundation for subsequent physical analyses.

\section{Geometrical acceptance effect}\label{GEO}

\begin{figure}
	\centering
	\includegraphics[width=.4\textwidth]{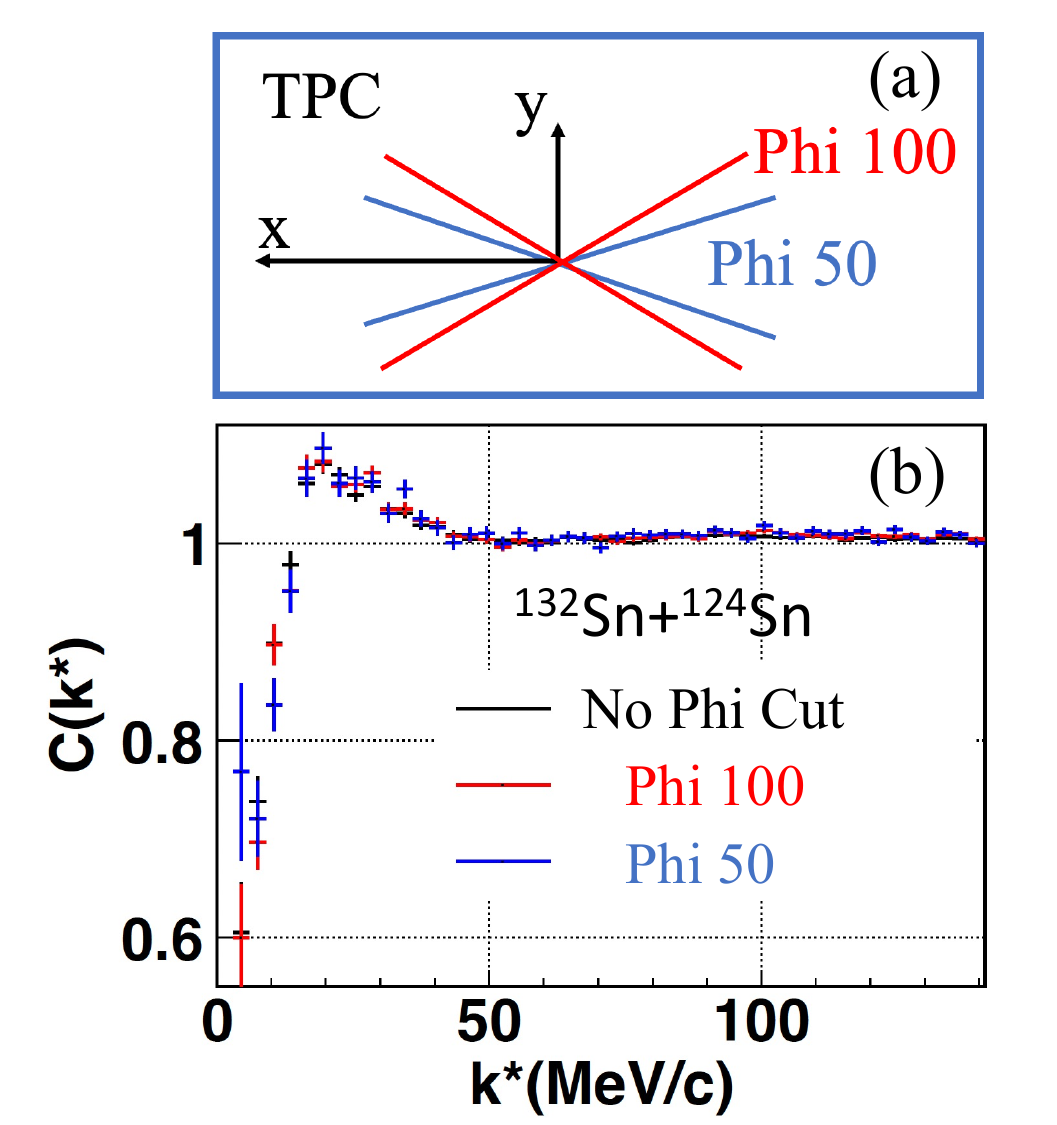}
	\caption{Schematic of different TPC angular acceptances (a) and the corresponding proton-proton correlation functions under various angular coverage conditions (b) for the $^{132}\text{Sn}+^{124}\text{Sn}$ reaction system. Taking Phi 100 as an example, it corresponds to an angular coverage of $\pm 50^{\circ}$ on both the positive and negative sides of the X-axis.}
	\label{fig8}
\end{figure}

The TPC detector employed in the S$\pi$RIT experiment features a rectangular structure with a relatively short dimension along the Y-axis direction. Consequently, particles emitted near the Y-axis direction in collisions leave short track lengths inside the TPC. In previous analyses of transverse momentum spectra, rapidity spectra, and collective flow \cite{SpiRIT:2021gtq,SpiRIT:2021och,SpiRIT:2022sqt,SpiRIT:2023htl}, a geometrical acceptance cut on the TPC angular coverage has already been adopted to suppress the inefficiency from short tracks.
To verify whether this angular selection affects the proton–proton correlation function, we investigate the variation of the proton–proton correlation function under different TPC angular acceptances. Figure \ref{fig8}(a) illustrates the schematic of different TPC angular coverage ranges (taking Phi 100 as an example, which corresponds to an angular range of $\pm 50^{\circ}$ on both positive and negative sides of the X-axis). Figure \ref{fig8}(b) compares the proton–proton correlation functions obtained with different angular acceptances for the $^{132}\text{Sn}+^{124}\text{Sn}$ reaction system.
It is evident that the shapes of the proton–proton correlation functions remain essentially consistent within the selected angular coverage range, indicating that the TPC angular acceptance has a negligible impact on the proton–proton correlation function under the present experimental conditions. To make full use of the experimental data, no additional angular acceptance cuts are applied in the subsequent analyses.

\section{Systematic uncertainty estimation}\label{SYS}

To evaluate the systematic uncertainty of the proton–proton correlation function, a detailed assessment is performed for several key selection criteria in the data analysis. Six critical selection cuts are chosen and varied systematically:(1) The distance $\texttt{ Dist}_{(\texttt{RAVE-POCA})}$ between the reaction vertex determined by the RAVE method \cite{Waltenberger:2011zz} and the point of closest approach (POCA) obtained from the Riemann circular arc fitting \cite{Fruhwirth:2018kke}; (2) The number of hit clusters used for track fitting, \texttt{ClusterNum}; (3) The proton mass parameter \texttt{massHCalib}; (4) The particle multiplicity threshold corresponding to the impact parameter, $\texttt{Multiplicity}$; (5) The track merging and splitting correction parameter $\Delta$\texttt{x} of the TPC in the XOZ plane; (6) The track merging and splitting correction parameter  $\Delta$\texttt{y} of the TPC along the Y-axis direction. By individually varying each of the above selection criteria and recalculating the proton–proton correlation function, we quantify the impact of each cut variation on the correlation function results.

\begin{figure*}
	\centering
	\includegraphics[width=.9\textwidth]{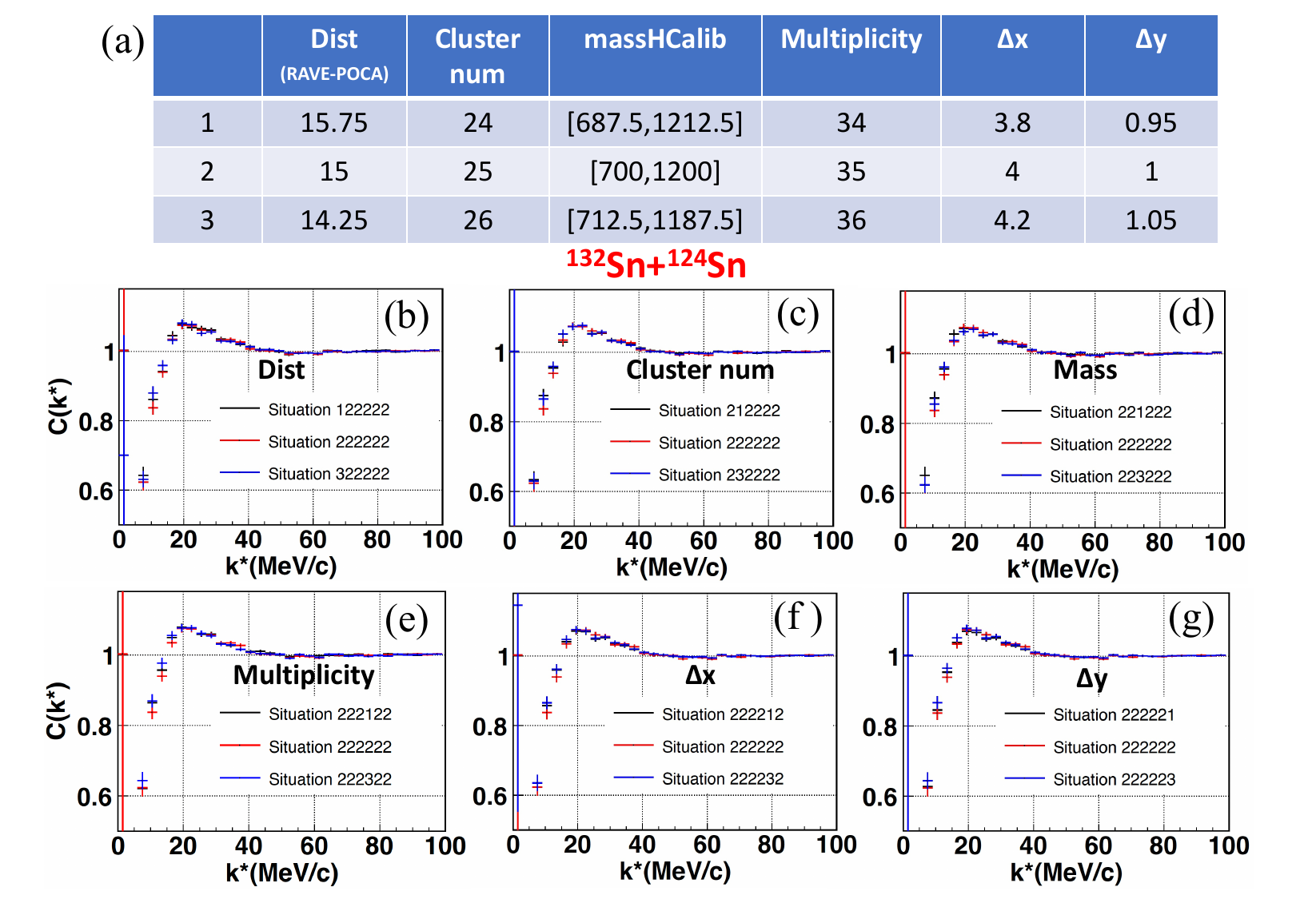}
	\caption{Systematic uncertainty analysis of the proton–proton correlation function for the $^{132}\text{Sn}+^{124}\text{Sn}$ reaction system. Panel (a) lists six selection criteria adopted for systematic uncertainty estimation together with their variation ranges; panels (b)–(g) illustrate the corresponding impacts on the proton–proton correlation function when each selection criterion is varied individually.}
	\label{fig9}
\end{figure*}

Figure \ref{fig9} presents the systematic uncertainty analyses of the proton–proton correlation functions for the $^{132}\text{Sn}+^{124}\text{Sn}$ reaction system. In the figure, panel (a) lists all selection criteria and their variation ranges, where the second row denotes the nominal values adopted in the standard analysis. { The first and third rows correspond to the varied values, respectively. The systematic uncertainty is evaluated based on the event selection properties of our experimental data and the performance characteristics of the S$\pi$RIT TPC. For integer‑valued cuts such as \texttt{ClusterNum} and \texttt{Multiplicity}, systematic uncertainties are estimated by shifting the cut value by $\pm 1$. This is because the number of clusters employed for reconstruction and particle multiplicity for centrality selection may introduce systematic biases in our results. For non-integer cuts including $\texttt{ Dist}_{(\texttt{RAVE-POCA})}$, $\Delta$\texttt{x} and $\Delta$\texttt{y}, systematic uncertainties are assessed by varying each cut by $\pm 5\%$, given that vertex position cuts and cuts on track pair separation may yield systematic effects; these varied values are consistent with the vertex and track resolutions \cite{SpiRIT:2020gce}. For the range cut \texttt{massHCalib}, since modifications to the selected mass window may introduce systematic effects, the variation is empirically set to $10\%$ of the full range width.} Panels (b)–(g) illustrate the corresponding influence on the proton–proton correlation function when each selection criterion is varied separately.
It is observed that variations of these selection criteria yield moderate effects on the proton–proton correlation function in the low relative-momentum region, while the overall distribution trend remains consistent.

{The total systematic uncertainty represents the combined contribution from individual sources. Since the uncertainties derived via the Monte Carlo method, which accounts for correlations among different parameters, are comparable to, or marginally smaller in some individual bins than those from the quadratic sum method, we adopt the quadratic sum method to yield a conservative estimate of the systematic uncertainty. At each relative momentum bin, the total systematic uncertainty is computed by summing in quadrature the systematic uncertainties originating from the six selection criteria.}

\begin{equation}
\sigma^{\text{sys}}_{\text{total}} = \sqrt{\sum_{i=1}^{6} \left( \sigma^{\text{sys}}_{i} \right)^{2}}
\label{eq_sys}
\end{equation}	

Here, $\sigma^{\text{sys}}_{\text{total}}$ denotes the total systematic uncertainty, and $\sigma^{\text{sys}}_{i}$ ($i=1,2,\dots,6$) corresponds to the systematic uncertainty component introduced by the individual variation of each of the six selection criteria mentioned earlier, namely $\texttt{ Dist}_{(\texttt{RAVE-POCA})}$, \texttt{ClusterNum}, \texttt{massHCalib}, $\texttt{Multiplicity}$, $\Delta \texttt{x}$ and $\Delta \texttt{y}$. 

\begin{figure}
	\centering
	\includegraphics[width=.4\textwidth]{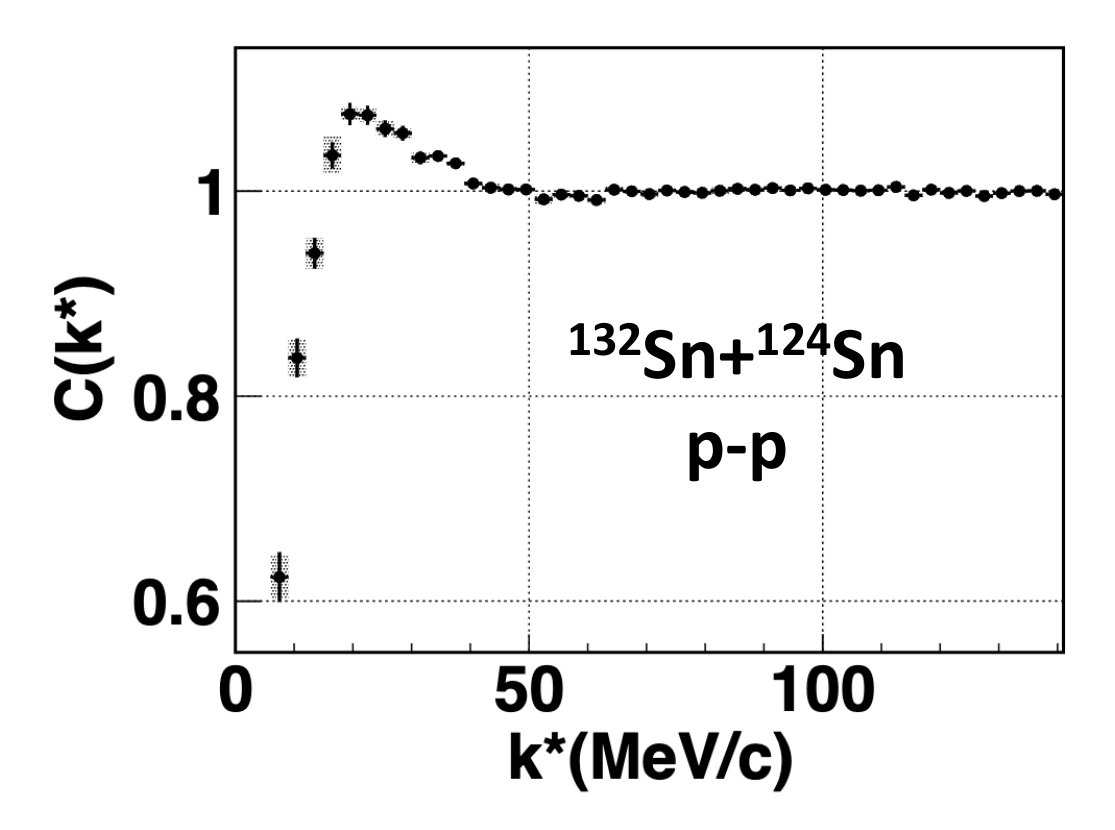}
	\caption{Experimental results of the proton–proton correlation function for the $^{132}\text{Sn}+^{124}\text{Sn}$ reaction system. The error bars on the data points represent the statistical uncertainties, and the shaded band denotes the systematic uncertainties.}
	\label{fig10}
\end{figure}

Figure \ref{fig10} represents the experimental results of the proton–proton correlation function for the $^{132}\text{Sn}+^{124}\text{Sn}$ reaction system. The data points represent the experimentally measured proton–proton correlation function values, the solid error bars denote the statistical uncertainties, and the shaded areas indicate the systematic uncertainties. It can be observed that the systematic uncertainties are relatively larger in the low relative-momentum region but remain controllable overall.
Furthermore, the proton–proton correlation function exhibits a positive correlation peak around 20~MeV/c, { originating from the attractive S‑wave nuclear force, while the anti‑correlation observed at small relative momenta is associated with the Pauli exclusion,  nucleon‑nucleon interaction, and Coulomb repulsion}. This data behavior is consistent with previous experimental results on proton–proton correlation functions \cite{Wang:2021mrv, STAR:2015kha}. These findings demonstrate that rectangular TPCs, such as the S$\pi$RIT TPC, installed in dipole magnets can successfully achieve the measurement of particle correlation functions in radioactive beam heavy-ion collisions. This work provides technical accumulation for future femtoscopy measurements in radioactive beam heavy-ion collisions using rectangular TPCs (e.g., the S$\pi$RIT TPC, CEE TPC \cite{Yang:2024sor}).

\section{Summary}\label{SUM}

Femtoscopy measurements with the S$\pi$RIT TPC are successfully implemented in radioactive beam heavy-ion collisions. A dedicated correction scheme for track merging and track splitting is proposed, which is well suited for rectangular TPCs deployed inside dipole magnets and effectively improves the quality of experimental correlation functions at small relative momenta.
Taking the proton–proton correlation function measured in the 270~{AMeV} $^{132}\text{Sn}+^{124}\text{Sn}$ radioactive beam heavy-ion collision system as an example, the correction of track merging and track splitting has been successfully applied, and the angular acceptance of the TPC exhibits a negligible influence on the proton–proton correlation function under the present experimental conditions. Furthermore, a systematic uncertainty quantification framework is established to evaluate the systematic errors of the correlation function, which provides a solid foundation for future high-precision studies of particle correlation functions in radioactive beam heavy-ion collisions.

\section{Acknowledgments}\label{Acknowledgments}

This experiment was performed at RI Beam Factory operated by RIKEN Nishina Center and CNS, University of Tokyo.
Computing resources were provided by the HOKUSAI-Great Wave system at RIKEN. 
This work was supported by the Nation Science Foundation of China under grant No.12205160 and 12335008, the Ministry of Science and Technology of China under Grant No. 2022YFE0103400, the Japanese MEXT KAKENHI
(Grant-in-Aid for Scientific Research on Innovative
Areas) grant No. 24105004, JSPS KAKENHI Nos.
JP17K05432 and JP19K14709, and JP21K03528, the U.S.
Department of Energy under Grant Nos. DE-SC0014530,
DE-NA0002923, and 
DE-FG02-93ER40773, the US National
Science Foundation Grant No. PHY-1565546, the
Polish National Science Center (NCN), Poland, under contract
Nos. UMO-2013/09/B/ST2/ 04064 and
UMO-2013/10/M/ST2/00624, and the National Research
Foundation of Korea (NRF) under Grant Nos.
2018R1A5A1025563, RS-2024-00333673 and RS-2024-00436392.
I. G. was supported by HIC for FAIR and the
Croatian Science Foundation under projects Nos. 1257 and 7194.
Z. G. Xiao acknowledges the support from the Initiative Scientific Research Program of Tsinghua University. 










\printcredits

\bibliographystyle{cas-model2-names}

\bibliography{cas-refs}



\end{document}

%% file: authors.tex
\author[1,2]{Y. J. Wang}
\cormark[1]
\cortext[1]{Corresponding author}
\ead{wyj25@buaa.edu.cn; Present address: School of Physics, Beihang University, Beijing 100191, China.}
\credit{ Writing original draft, Visualization, Validation, Investigation, Formal analysis,  Conceptualization, Supervision, Funding acquisition. }
\author[3]{C. K. Tam}
\credit{Investigation, Conceptualization.}
\author[1,4]{Z. G. Xiao}
\cormark[1]
\ead{xiaozg@tsinghua.edu.cn}
\credit{Writing original draft, Validation, Investigation, Conceptualization, Supervision, Funding acquisition.}
\author[5,6]{W. G. Lynch}
\cormark[1]
\ead{lynch@nscl.msu.edu}
\credit{Validation, Investigation, Conceptualization, Supervision, Funding acquisition.}
\author[5,6]{C. Y. Tsang}
\credit{Investigation}
\author[5,6]{J. Barney}
\credit{Investigation}
\author[5]{G. Jhang}
\credit{Investigation}
\author[5,6]{J. Estee}
\credit{Investigation}
\author[5,6]{M. B. Tsang}
\cormark[1]
\ead{tsang@nscl.msu.edu}
\credit{Validation, Investigation, Conceptualization, Supervision, Funding acquisition.}
\author[7]{R. Wang}
\credit{Investigation}
\author[2,8]{M. Kaneko}
\credit{Investigation}
\author[9]{J. W. Lee}
\credit{Investigation}
\author[9]{J. Park}
\credit{Investigation}
\author[3]{Z. Chajęcki}
\cormark[1]
\ead{zbigniew.chajecki@wmich.edu}
\credit{Validation, Investigation, Conceptualization, Supervision.}
\author[10]{G. Verde}
\cormark[1]
\ead{giuseppe.verde@ct.infn.it}
\credit{Validation, Investigation, Conceptualization, Supervision.}
\author[2]{T. Isobe}
\cormark[1]
\ead{isobe@riken.jp}
\credit{Validation, Investigation, Conceptualization, Supervision, Funding acquisition.}
\author[2]{M. Kurata-Nishimura}
\credit{Investigation}
\author[2,8]{T. Murakami}
\credit{Investigation}
\author[2]{D. S. Ahn}
\credit{Investigation}
\author[11,12]{L. Atar}
\credit{Investigation}
\author[11,12]{T. Aumann}
\credit{Investigation}
\author[2]{H. Baba}
\credit{Investigation}
\author[12]{K. Boretzky}
\credit{Investigation}
\author[13]{J. Brzychczyk}
\credit{Investigation}
\author[5]{G. Cerizza}
\credit{Investigation}
\author[2]{N. Chiga}
\credit{Investigation}
\author[2]{N. Fukuda}
\credit{Investigation}
\author[14,2,11]{I. Gasparic}
\credit{Investigation}
\author[9]{B. Hong}
\credit{Investigation}
\author[11,12]{A. Horvat}
\credit{Investigation}
\author[15]{K. Ieki}
\credit{Investigation}
\author[2]{N. Inabe}
\credit{Investigation}
\author[16]{Y. J. Kim}
\credit{Investigation}
\author[17]{T. Kobayashi}
\credit{Investigation}
\author[18]{Y. Kondo}
\credit{Investigation}
\author[13]{P. Lasko}
\credit{Investigation}
\author[16]{H. S. Lee}
\credit{Investigation}
\author[12]{Y. Leifels}
\credit{Investigation}
\author[19]{J. Łukasik}
\credit{Investigation}
\author[5,6]{J. Manfredi}
\credit{Investigation}
\author[20]{A. B. McIntosh}
\credit{Investigation}
\author[5]{P. Morfouace}
\credit{Investigation}
\author[18]{T. Nakamura}
\credit{Investigation}
\author[2,8]{N. Nakatsuka}
\credit{Investigation}
\author[2]{S. Nishimura}
\credit{Investigation}
\author[2]{H. Otsu}
\credit{Investigation}
\author[19]{P. Pawłowski}
\credit{Investigation}
\author[13]{K. Pelczar}
\credit{Investigation}
\author[11]{D. Rossi}
\credit{Investigation}
\author[2]{H. Sakurai}
\credit{Investigation}
\author[5]{C. Santamaria}
\credit{Investigation}
\author[2]{H. Sato}
\credit{Investigation}
\author[11]{H. Scheit}
\credit{Investigation}
\author[5]{R. Shane}
\credit{Investigation}
\author[2]{Y. Shimizu}
\credit{Investigation}
\author[12]{H. Simon}
\credit{Investigation}
\author[21]{A. Snoch}
\credit{Investigation}
\author[13]{A. Sochocka}
\credit{Investigation}
\author[2]{T. Sumikama}
\credit{Investigation}
\author[2]{H. Suzuki}
\credit{Investigation}
\author[2]{D. Suzuki}
\credit{Investigation}
\author[2]{H. Takeda}
\credit{Investigation}
\author[22,23]{S. Tangwancharoen}
\credit{Investigation}
\author[11,12]{H. Toernqvist}
\credit{Investigation}
\author[15]{Y. Togano}
\credit{Investigation}
\author[20,24]{S. J. Yennello}
\credit{Investigation}
\author[25]{Y. Zhang}
\credit{Investigation}
\author{the S$\pi$RIT collaboration}
\affiliation[1]{
  organization={Department of Physics, Tsinghua University},
  city={Beijing},
  postcode={100084},
  country={China}
}
\affiliation[2]{
  organization={RIKEN Nishina Center},
  addressline={Hirosawa 2-1},
  city={Wako, Saitama},
  postcode={351-0198},
  country={Japan}
}
\affiliation[3]{
  organization={Department of Physics, Western Michigan University},
  city={Kalamazoo, Michigan},
  postcode={49008},
  country={USA}
}
\affiliation[4]{
  organization={Center for High Energy Physics, Tsinghua University},
  city={Beijing},
  postcode={100084},
  country={China}
}
\affiliation[5]{
  organization={National Superconducting Cyclotron Laboratory, Michigan State University},
  city={East Lansing, Michigan},
  postcode={48824},
  country={USA}
}
\affiliation[6]{
  organization={Department of Physics, Michigan State University},
  city={East Lansing, Michigan},
  postcode={48824},
  country={USA}
}
\affiliation[7]{
  organization={School of Radiation Medicine and Protection, Soochow University},
  city={Suzhou},
  country={China}
}
\affiliation[8]{
  organization={Department of Physics, Kyoto University},
  addressline={Kita-shirakawa},
  city={Kyoto},
  postcode={606-8502},
  country={Japan}
}
\affiliation[9]{
  organization={Department of Physics, Korea University},
  city={Seoul},
  postcode={02841},
  country={Republic of Korea}
}
\affiliation[10]{
  organization={INFN - Sezione di Catania},
  city={Catania},
  postcode={95123},
  country={Italy}
}
\affiliation[11]{
  organization={Institut für Kernphysik, Technische Universität Darmstadt},
  city={Darmstadt},
  postcode={D-64289},
  country={Germany}
}
\affiliation[12]{
  organization={GSI Helmholtzzentrum für Schwerionenforschung},
  addressline={Planckstrasse 1},
  city={Darmstadt},
  postcode={64291},
  country={Germany}
}
\affiliation[13]{
  organization={Faculty of Physics, Astronomy and Applied Computer Science, Jagiellonian University},
  city={Kraków},
  country={Poland}
}
\affiliation[14]{
  organization={Division of Experimental Physics, Rudjer Boskovic Institute},
  city={Zagreb},
  country={Croatia}
}
\affiliation[15]{
  organization={Department of Physics, Rikkyo University},
  addressline={Nishi-Ikebukuro 3-34-1},
  city={Tokyo},
  postcode={171-8501},
  country={Japan}
}
\affiliation[16]{
  organization={Rare Isotope Science Project, Institute for Basic Science},
  city={Daejeon},
  postcode={34047},
  country={Republic of Korea}
}
\affiliation[17]{
  organization={Department of Physics, Tohoku University},
  city={Sendai},
  postcode={980-8578},
  country={Japan}
}
\affiliation[18]{
  organization={Department of Physics, Tokyo Institute of Technology},
  city={Tokyo},
  postcode={152-8551},
  country={Japan}
}
\affiliation[19]{
  organization={Institute of Nuclear Physics PAN},
  addressline={ul. Radzikowskiego 152},
  city={Kraków},
  postcode={31-342},
  country={Poland}
}
\affiliation[20]{
  organization={Cyclotron Institute, Texas A\&M University},
  city={College Station, Texas},
  postcode={77843},
  country={USA}
}
\affiliation[21]{
  organization={Nikhef National Institute for Subatomic Physics},
  city={Amsterdam},
  country={Netherlands}
}
\affiliation[22]{
  organization={Department of Physics, Faculty of Science, King Mongkut's University of Technology Thonburi},
  city={Bangkok},
  postcode={10140},
  country={Thailand}
}
\affiliation[23]{
  organization={Center of Excellence in Theoretical and Computational Science (TACS-CoE), Faculty of Science, King Mongkut's University of Technology Thonburi},
  city={Bangkok},
  postcode={10140},
  country={Thailand}
}
\affiliation[24]{
  organization={Department of Chemistry, Texas A\&M University},
  city={College Station, Texas},
  postcode={77843},
  country={USA}
}
\affiliation[25]{
  organization={School of Physics and Technology, Nanjing Normal University},
  city={Nanjing},
  postcode={210023},
  country={China}
}